\documentstyle[aps,epsfig,epsf]{revtex}
\def\lsim{\raise0.3ex\hbox{$<$\kern-0.75em\raise-1.1ex\hbox{$\sim$}}}
\def\gsim{\raise0.3ex\hbox{$>$\kern-0.75em\raise-1.1ex\hbox{$\sim$}}}
\def\be{\begin{equation}}
\def\ee{\end{equation}}
\def\ba{\begin{eqnarray}}
\def\ea{\end{eqnarray}}
\def\bea{\begin{eqnarray}}
\def\eea{\end{eqnarray}}
 2
\def\m0{m_{D0}}
\def\bx{{\bf x}}

\def\half{{1\over 2}}

\newcommand{\tr}{{\rm Tr}}
\def\d{\partial}

\begin{document}
\title{Propagators and Dimensional Reduction of Hot SU(2) Gauge Theory}
\author{
A. Cucchieri}
\address{IFSC-USP,
Caixa Postal 369, 13560-970 Sao Carlos, SP, Brazil}
\author{F. Karsch and P. Petreczky}
\address{
Fakult\"at f\"ur Physik, Universit\"at Bielefeld,\\
\qquad  P.O. Box 100131, D-33501 Bielefeld, Germany\\
}
\date{\today}
\maketitle
\vspace*{-5.0cm}
\noindent
\hfill \mbox{BI-TP 2001/04}
\vspace*{4.8cm}
\begin{abstract}
We investigate the large distance behavior of the electric 
and magnetic propagators of hot SU(2) gauge theory in different gauges
using lattice simulations of the full 4d theory and the effective,
dimensionally reduced 3d theory. Comparison of the 3d and 4d 
propagators suggests that dimensional reduction works surprisingly well
down to the temperature $T=2 T_c$. Within statistical uncertainty
the electric screening mass is found to be gauge independent.
The magnetic propagator, on the other hand, 
exhibits a complicated gauge dependent structure at low
momentum.
\end{abstract}
\section{Introduction} 
Understanding the large distance behavior of the static electric and magnetic 
propagators in hot SU(N) gauge theory is of interest for several reasons.
First of all it is related to the screening phenomenon in a hot
non-Abelian plasma. In fact, the concepts of electric and magnetic masses
extracted from these propagators form to a large extent the basis for our intuitive
understanding of screening in non-abelian gauge theories \cite{Gross81}.
Furthermore, the concept of screened electric propagators finds application 
in refined perturbative calculations (hard thermal loop resummation)
\cite{karsch97,andersen99,blaizot99}. 
Non-perturbative
calculations of these propagators thus can provide a bridge between perturbative and 
non-perturbative descriptions of the electric screening phenomenon.

Of course, a draw back of such an approach is that the gluon propagator itself
is a gauge dependent quantity. One thus has to question to what extent the 
results extracted from these propagators have a physical meaning.
The poles of the gluon propagator at finite temperature were
proven to be gauge invariant in perturbation theory \cite{kobes90}.
However, for static quantities like the Debye mass static magnetic fields
can lead to a breakdown of perturbation theory. The Debye mass cannot be
defined perturbatively  beyond leading order \cite{rebhan94,arnold95}.
Non-perturbatively the problem of gauge 
independence of the screening masses thus is an open question.

It has been shown that at high temperature the large distance behavior of a 
SU(N) gauge theory
can be described in terms of the dimensionally reduced effective theory, i.e. the
three dimensional 
adjoint Higgs model 
\cite{karsch98,lacock92,karkkainen93,karkkainen94,kajantie97a,laine99,hart99,hart00}.
The screening masses extracted from gauge invariant correlators were
studied in terms of the effective three dimensional (3d) theory 
\cite{kajantie97a,laine99,hart99,hart00,kajantie97b,laine98} and were compared with
the corresponding results from four dimensional (4d) simulations \cite{datta98,datta99}.
The relation between the propagator
masses and the masses extracted from gauge invariant correlators was discussed in
\cite{karsch98}.
The screening masses extracted from gauge invariant correlators correspond
to masses of some bound states of the 3d effective theory and are several times
larger than masses extracted from propagators. 

In the present paper we  extend our earlier studies on the electric
and magnetic propagators \cite{karsch98,heller95,heller98}. Contrary to
Refs. \cite{karsch98,heller95,heller98} where propagators were studied only
in Landau gauge we  consider here a class of generalized Landau gauges,
the Coulomb gauge, the maximally Abelian gauge and the static time averaged
Landau gauge.
We study the
propagators  in terms of the 3d effective theory as well as  in the 
4d theory at finite temperature.
A comparison of the propagators in the full 4d and in the 3d effective
theory  provides further evidence for the applicability of 
dimensional reduction. A detailed 
study of finite size effects in the propagators shows that
the picture of  magnetic screening given in 
Refs. \cite{karsch98,heller95,heller98} needs to be revised to some extent.
First results from our analysis of the magnetic sector
have been presented in Ref. \cite{plb}.

The paper is organized as  follows. In section II we 
discuss questions related to gauge fixing and the definition of
propagators in finite temperature SU(2) theory as well as  in the dimensionally
reduced effective theory, the 3d adjoint Higgs model. 
Section III contains our
main results on electric and magnetic propagators and the analysis of their gauge
and volume dependence. Finally we give  our conclusion in section IV. In an Appendix
details of the determination of the parameters of the effective
theory are discussed. 

\section{Gluon propagators in finite temperature SU(2) gauge theory}

In this section we  define the actions we use for our 
simulations in three and four dimensions and introduce our basic notation
for gauge fields, gluon propagators
and the different gauges we have analyzed.

\subsection{Actions in three and four dimensions}

In four dimensions (4d) all our calculations are performed with the standard
Wilson action for SU(2) lattice gauge theory,
\be
S_{W}=\beta_4 \sum_{x, \nu>\mu} \bigl[ 1 - \half \tr U_\mu(x)
U_\nu(x+\hat{\mu})U_\mu^\dagger(x+\hat{\nu})U_\nu^\dagger(x)  \bigr],
\ee
where $U_\mu(x)\in SU(2)$ are the usual link variables and 
$\beta_4={4 /g_4^2}$. In three
dimensions (3d) the standard dimensional reduction process leads us to 
consider the 3d adjoint Higgs model
\bea 
S&=&-\beta_3 \sum_{\bx, \nu > \mu} \half \tr 
U_\mu(\bx)
U_\nu(\bx+\hat{\mu})U_\mu^\dagger(\bx+\hat{\nu})U_\nu^\dagger(\bx)
-\beta_3 \sum_{\bx,\hat \mu} \half \tr
A_0(\bx) U_\mu(\bx) A_0(\bx+\hat \mu)
U_\mu^{\dagger}(\bx) \nonumber\\
&&\quad
+\beta_3 \sum_{\bx} \left[\left(3+\half h\right) \half \tr A_0^2(\bx) +
x { \left( \half \tr
A_0^2(\bx)\right)}^2 \right],
\label{act}
\eea
where $\beta_3$ now is related to the dimensionful 3d gauge coupling and the 
lattice spacing $a$, {\it i.e.}  $\beta_3={4 / g_3^2a}$. The adjoint Higgs 
field is parameterized 
by hermitian matrices $A_0=\sum_a \sigma^a
A_0^a$ ($\sigma^a$ {are the usual Pauli matrices}) \cite{kajantie97a}.
Furthermore, $x$ parameterizes the quartic self coupling of the Higgs field 
and $h$ denotes the bare Higgs mass squared. 
The relation between the 3d and
the 4d couplings  will be discussed in an Appendix.
We also note that the indices $\mu,~\nu$, of course, run from 0 to 3 in  four
dimensions and from $1$ to $3$ in three dimensions. Although we will not always
mention this difference explicitly it should be obvious from the context how
various sums that appear have to be interpreted.  

\subsection{Gauge fixing}

As we want to analyze properties of the gluon propagator, which is a gauge dependent 
quantity, we have to fix a gauge on each configuration on which we want to calculate
this observable. In the past most studies of the gluon propagator have been performed
in Landau gauge. Here we will consider a class of $\lambda$-gauges which are 
generalizations of gauges that have been introduced in \cite{bernard90,bernard91} to
smoothly interpolate between Landau and Coulomb gauge. In the continuum these gauges 
correspond to the gauge condition
\be
\sum_{\mu} \lambda_\mu \d_\mu A_\mu =0 \quad.
\ee
On the lattice these gauges are realized by maximizing the quantity
\be
\tr \sum_{\mu,x} \lambda_{\mu} U_{\mu}(x) \quad.
\label{gfix}
\ee
In 4d the Landau gauge condition thus corresponds to the case $\lambda_\mu\equiv 1$ for all 
$\mu=0,..,3$, while the Coulomb gauge is given by 
$\lambda_0 =0$ and $\lambda_i=1$ for $i=1,~2,~3$. 
In the latter case we have to impose an additional gauge condition for the residual gauge
degree of freedom. We do this by demanding 
\be
\sum_{\bx} U_0(x_0,\bx)=u_0 \quad,
\label{Coulombadd}
\ee
to be independent of $x_0$.
In addition to the $\lambda$-gauges we also consider the Maximally Abelian gauge
(MAG) which can be realized by maximizing the quantity \cite{kronfeld87}
\be
\sum_{x,\mu} \tr \bigl[\sigma_3 U_{\mu}(x) \sigma_3 U_{\mu}^{\dagger}(x)\bigr] \quad,
\ee
with $\sigma_3$ being a  Pauli matrix.
Also in this case one has to fix a residual gauge degree of freedom which we do by
imposing a  U(1)-Landau gauge condition \cite{amemiya99}. 
In 4d SU(2) gauge theory we also consider the static time-averaged Landau gauge
(STALG) introduced in \cite{curci85,reisz91}. In the continuum it is defined by
\be
\d_0 A_0(x_0,\bx)=0,~~~~\sum_{x_0} \sum_{i=1}^3 \d_{i} A_{i}=0.
\ee
On lattice this gauge is realized in two steps. First we  maximize the
quantity $\tr \sum_x U_0(x)$.
In the second step we maximize the quantity
\be
\tr \sum_{i=1}^3 \sum_x U_i(x)
\ee
performing $x_0$ -independent gauge transformations.

While the notion of Landau and Maximally Abelian gauges carries over easily to the 3d case
we have to explain a bit more in detail our notion of $\lambda$-gauges in 3d. We have 
considered two versions of $\lambda$-gauge:
\begin{itemize}
\item[ ]
\be
\lambda_3{\rm -gauge:}\quad \d_1  A_1 + \d_2  A_2 +\lambda_3  \d_3  A_3=0 \quad
\ee
\item[ ]
\be
\lambda_1{\rm -gauge:}\quad \lambda_1 \d_1 A_1 + \d_2  A_2 +\d_3  A_3=0 \quad
\ee
\end{itemize}
The $\lambda_1$-gauges are more closely related to the 4d $\lambda$-gauges considered by us;
in both cases the rotational symmetry of the gauge condition is broken in a direction
orthogonal to the $x_3$ (or $z$) direction which we are going to use for separating the
sources in our correlation functions. Furthermore, we introduce in 3d the
so-called
Coulomb gauges ($c_1$ gauges)
which fix the gauge in a plane transverse to the $z$-direction,

\begin{itemize}
\item[ ]
\be 
c_1{\rm -gauge:}\quad c_1 \d_1 A_1 + \d_2  A_2  =0 \quad . \qquad\qquad
\ee
\end{itemize}
Of course, as in the 4d case 
we again need an additional gauge condition for the residual gauge degree of freedom.
The case $c_1=1.0$ is the usual Coulomb gauge.
This gauge has the advantage that a positive definite transfer matrix exists in the z-direction.

In our numerical calculations the gauge fixing was performed using a standard 
iterative algorithm accelerated by an overrelaxation step \cite{mandula90} as well 
as by using a stochastic overrelaxation algorithm \cite{forcrand89}. 

When analyzing observables in a fixed gauge one also has to address
the question to what extent Gribov copies can influence the result.
This problem will not be discussed here. It previously has been studied in 
4d SU(2) gauge theories 
at zero temperature \cite{cucchieri97} as well as at non-zero temperature
\cite{heller95}. In both cases no influence of Gribov
copies on the gluon propagators was found within the statistical accuracy
achieved in these studies.

\subsection{Gauge fields and gluon propagators} 

Definitions of the lattice gauge field $A_{\mu}(x)$ utilize the naive continuum 
relation between the lattice link variables $U_{\mu}(x)$ and the gauge field
variables in the continuum limit, {\it i.e.} 
$U_{\mu}(x) = \exp{(i g a A_{\mu}(x))}$. 
A straightforward definition thus is 
\ba
A_{\mu}(x) & = &{1 \over 2\, i g a}\,
\left[\,U_{\mu}(x)\,-\,U_{\mu}^{\dagger}(x)\,\right]
\;\mbox{,}
\label{eq:Amux}
\ea
which in the continuum limit differs from the continuum gauge fields by
${\cal O}((ag)^2)$ corrections.  
Here $a$ is the lattice spacing and $g$ either the 3d or 4d bare gauge coupling.
Other possible definitions 
for $A_{\mu}(x)$ which formally lead to smaller discretization errors 
were considered in \cite{cucchieri00proc}. It was found there 
that up to an overall multiplicative constant
the difference in magnetic and electric propagators due to different
definitions
of the gauge field is much smaller than the statistical errors.
We note that the definition (\ref{eq:Amux}) of the gauge field assumes that
the link variable $U_{\mu}$ is close to the unit matrix.
This seems to be the case for all gauges considered except 
the Coulomb gauge. In the case of Coulomb gauge the gauge fixing
procedure does not force the temporal link to be close to the unit
matrix. 
Only the spatial links turn out to be close to the unit matrix.
For instance, for links averaged over a  lattice volume of size 
$12^2 \times 24 \times 4$ we find for $\beta_4=2.512$ (corresponding to
$2 T_c$)
$< \tr U_0/2 > = 0.576(4)$ and $< \tr U_i/2 > = 0.90089(6)$.
These numbers should be compared with the corresponding ones in Landau gauge:
$< \tr U_0/2 > = 0.8879(1)$ and $<  \tr U_i/2 > = 0.88709(7)$.
One can expect that $\tr U_0/2$ gets closer to 1 as one approaches  
the continuum limit. Indeed for the Coulomb gauge on a $24^2 \times 48 \times 8$
lattice at $\beta=2.74$ (also corresponding to $2 T_c$)
we find  $< \tr U_0/2 > = 0.625(6)$. However, from these numbers
it is clear that very large values of $N_{\tau}$, i.e. large couplings $\beta_4$,
are necessary to get a
meaningful definition of the $A_0$ field. Therefore we will use the
4d Coulomb gauge only to analyze magnetic propagators.

The lattice gluon propagators in d-dimension are defined as 
\ba
D_{\mu \mu}(z) & = & 
\frac{1}{a^d \Omega}
\biggl\langle a \sum_{x_3, b} \,
Q^b_{\mu}(x_3 + z)\, Q^b_{\mu}(x_3)\,\biggr\rangle\;\mbox{.}
\ea
Here $Q^b_{\mu}(x_3)$ 
is a sum over all gauge fields in a hyperplane orthogonal to $x_3$,  
\ba
Q^b_{\mu}(x_3) & = & a^{d-1}  \sum_{x_{\perp}} \, A^b_{\mu}(x_{\perp}\mbox{,}\,x_3)
\;\mbox{,}
\ea
with $A^b_{\mu}(x)={1\over 2} \tr [A_{\mu}(x)\sigma^b]$; 
$x_{\perp}= (x_1,x_2)$ in 3d and 
$x_{\perp}= (x_0,x_1,x_2)$ in 4d, respectively.
Furthermore, $\Omega=N^2N_z$ is the lattice volume in 3d and $\Omega=N^2 N_z
N_t$ in 4d \footnote{We will also use the notation $V=N^2 N_z$ for the spatial
volume in 3d and 4d.}.
The electric and magnetic propagators are then defined in the usual
way (see e.g. Refs. \cite{heller95,heller98}),
\be
D_E(z)=D_{00}(z)\quad , \quad D_M(z) =  {1\over 2}\left( D_{11}(z)+D_{22}(z) \right)  \quad.
\ee
$D_{33}(z)$ is not included in the definition of $D_M(z)$ because it is constant
in $\lambda$ and Coulomb gauges.
We will also consider momentum space propagators which are obtained through a one dimensional
Fourier transformation, 
\be
\tilde{D}_{\mu \mu}(k)=a \sum_{z=0}^{N_z-1} \, e^{i k z} D_{\mu \mu}(z),
\mbox{with}~~~ k={2 \pi n\over N_z},~n=0,1,....N_z/2 
\label{fmu}
\ee
We use the standard definition for the momentum space magnetic propagator
(see e.g. \cite{cucchieri99,bernard94})
\be
\tilde D_M(k)=\frac{1}{3} \sum_{\mu=1}^3 D_{\mu \mu}(k).
\ee
We note that we include here $\tilde D_{33}$ in the definition of the momentum
space propagators in order to take into account the contribution of the
constant component $D_{33}(z)$, which only influences the zero mode contribution
$\tilde D_{\mu \mu}(k=0)$.
The electric propagator in momentum space is simply defined by 
$\tilde D_E(k)=\tilde D_{00}(k)$.
In order to absorb additional cut-off effects in momentum space propagators we
find it usefull to analyze these in terms of the momenta $p \equiv |2 \sin(k/2)|$
rather than the lattice momenta $k$. In the following we thus will use $p$
instead of $k$ as a definition for our lattice momenta.
For the analysis of the long distance behavior of these correlation functions it is customary
and, in fact, quite instructive to consider {\it local masses}, which are defined by 
\be
{D_i(z)\over D_i(z+1)}={\cosh(m_i(z) (z-N_z/2))\over \cosh(m_i(z)(z+1-N_z/2))},
\label{localm}
\ee
with $i=E,M$. If the propagators decay exponentially starting from some value of $z$, the 
corresponding local masses will reach a plateau. 

All our 4d simulations have been performed at a temperature $T=2 T_c$. The values for 
the 4d coupling
$\beta_4$ corresponding to this temperature were taken from Ref. \cite{heller98}
and are given in Table I together with the corresponding lattice volumes used in our 4d simulations.
\vskip 0.5truecm
\begin{center}
\begin{tabular}{|c|c|}
\hline
$~~\beta_4=2.512~~$ & $~~~8^2 \times 16 \times 4~~$ \\
$~~~~~~~~~~~~~~~$ & $~~12^2 \times 24 \times 4~~$ \\
$~~~~~~~~~~~~~~~$ & $~~16^2 \times 32 \times 4~~$ \\
$~~~~~~~~~~~~~~~$ & $~~20^2 \times 40 \times 4~~$ \\
$~~~~~~~~~~~~~~~$ & $~~24^2 \times 48 \times 4~~$ \\
$~~~~~~~~~~~~~~~$ & $~~28^2 \times 56 \times 4~~$ \\
$~~~~~~~~~~~~~~~$ & $~~32^2 \times 64 \times 4~~$ \\
\hline
$~~\beta_4=2.740~~$ & $~~16^2 \times 32 \times 8~~$ \\
$~~~~~~~~~~~~~~~$ & $~~24^2 \times 48 \times 8~~$ \\
\hline
\end{tabular}

\vskip0.5truecm
TABLE I. Couplings and lattice volumes used in the 4d simulations
of finite temperature SU(2) gauge theory.
\end{center}
\vskip0.5truecm

We  give the choice of parameters for our 3d simulations in the next
subsection. The cutoff dependence is 
discussed for the magnetic and electric propagators separately in section III.

\subsection{3d versus 4d calculations}

Most of the results we are going to discuss in the following section have been
obtained through simulations in the  dimensionally reduced version of the
4d SU(2) gauge theory at finite temperature. As we do want to compare our 
results obtained in 3d with corresponding results in the 4d theory obtained
at a temperature $T=2 T_c$ we should check that dimensional reduction yields
reliable results at temperatures this close to the critical point. 
In Refs. \cite{laine99,hart99,hart00} it was shown that the effective theory is capable
to describe the long distance behavior of some gauge invariant correlators
in SU(2) and SU(3) gauge theories at these temperatures.
A similar conclusion has been reached
for the 2d dimensionally reduced version of the 3d SU(3) gauge theory at finite 
temperature \cite{bialas00}. 

\begin{figure}
\epsfxsize=6cm
\epsfysize=6cm
\centerline{\epsffile{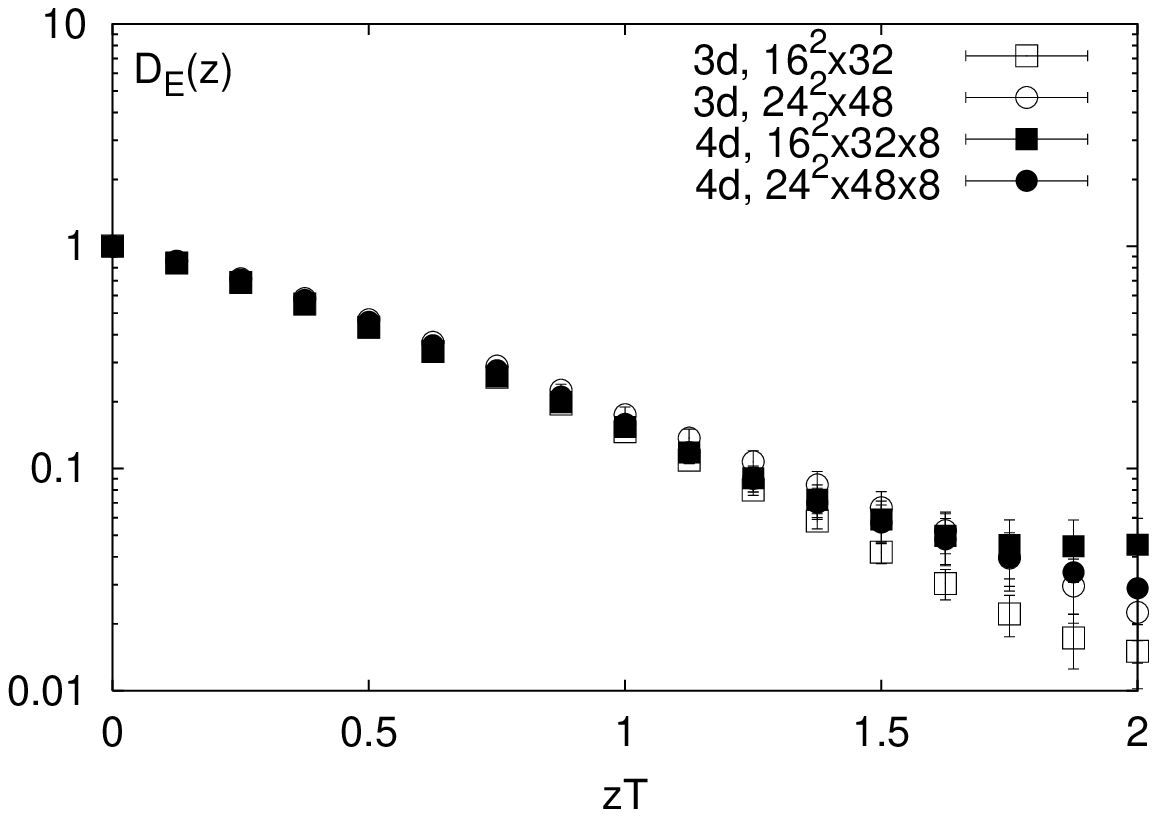}
\epsfxsize=6cm \epsfysize=6cm \epsffile{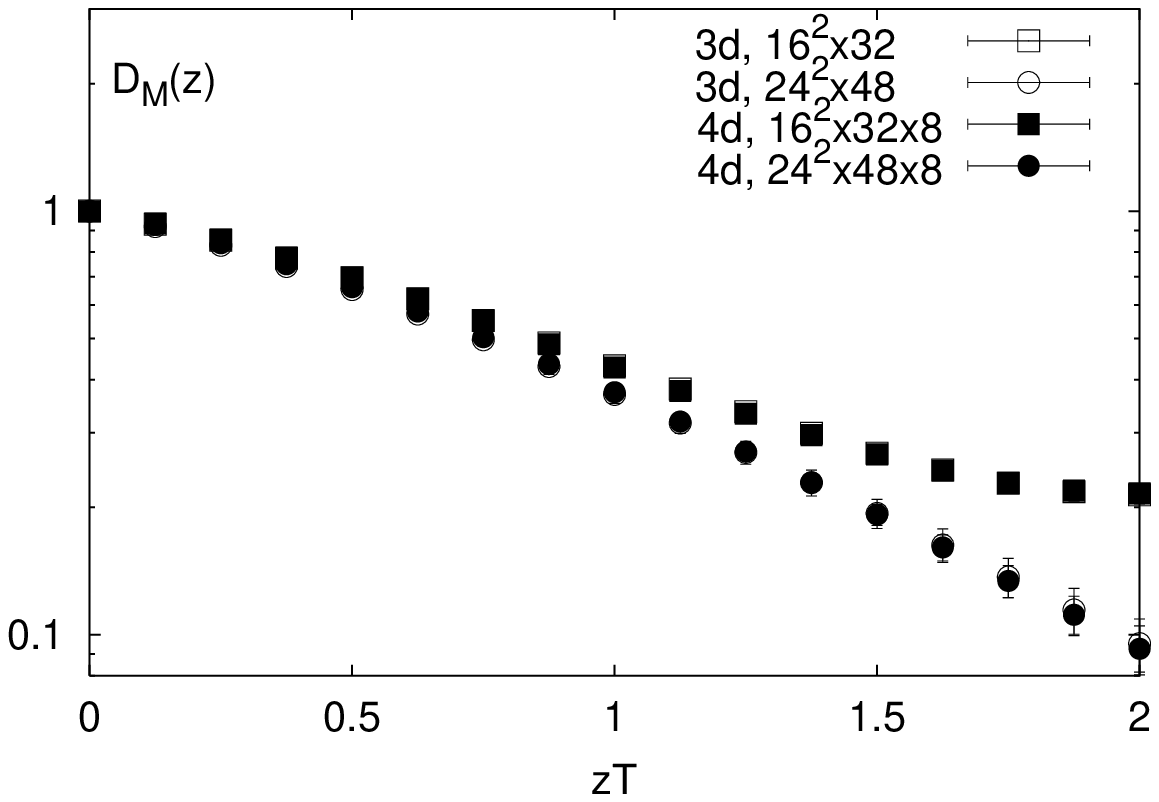} }
\caption{Comparison of 4d and 3d data for the electric (left) and magnetic
(right) gluon propagator in Landau gauge calculated at $T= 2 T_c$.
}
\label{match}
\end{figure}

\begin{figure}
\epsfxsize=6cm
\epsfysize=6cm
\centerline{\epsffile{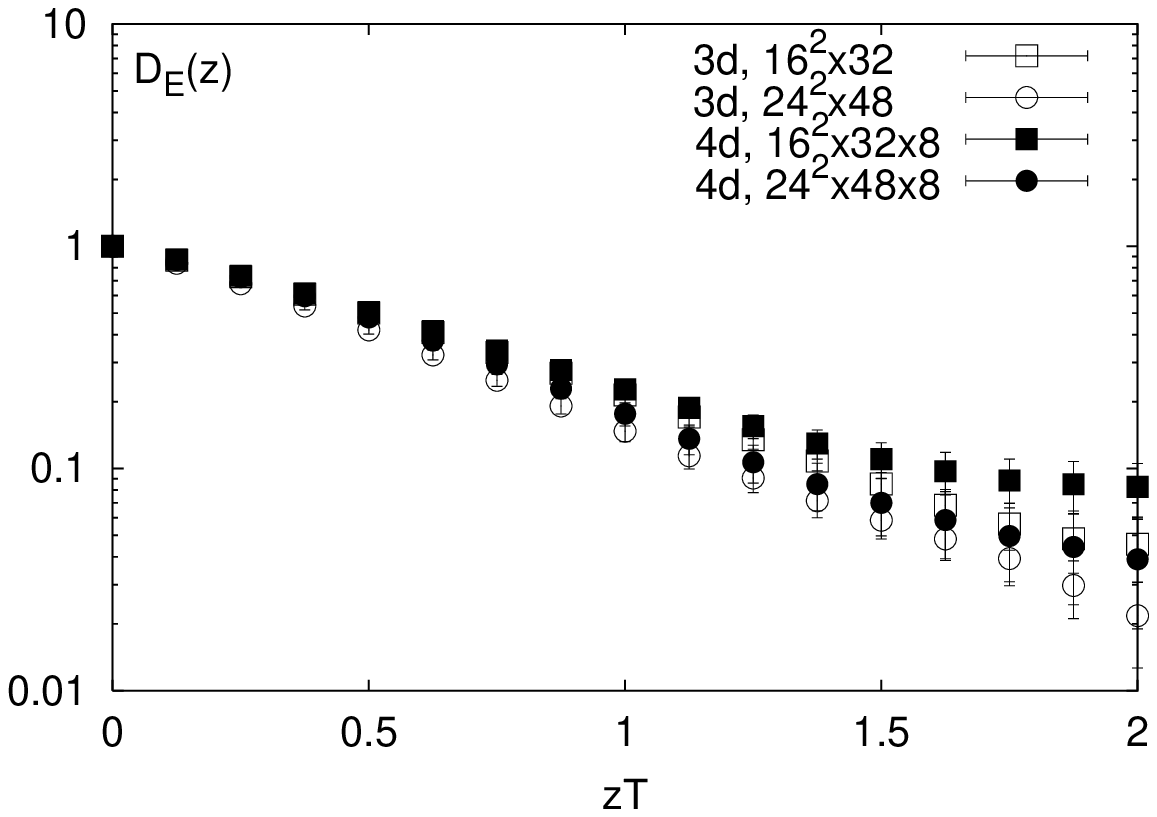}
\epsfxsize=6cm \epsfysize=6cm \epsffile{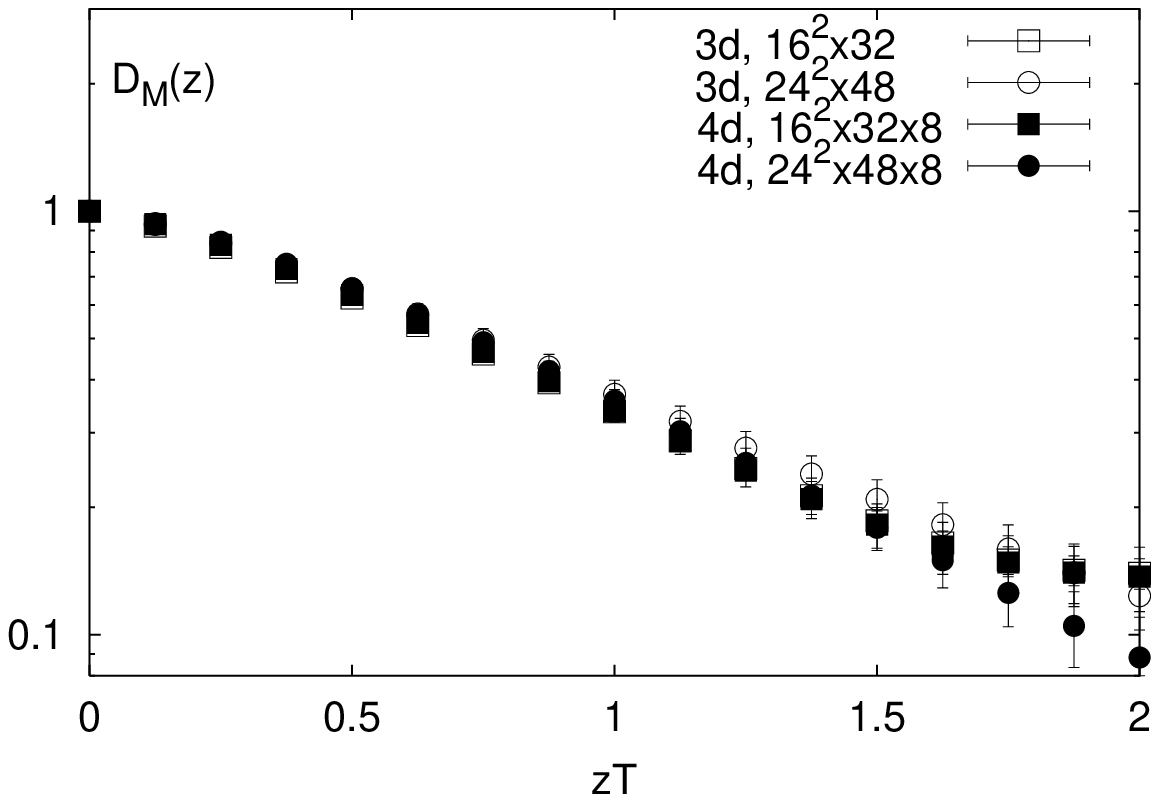} }
\caption{Comparison of 4d and 3d data for the electric (left) and magnetic
(right) gluon propagator in Maximally Abelian gauge calculated at $T= 2 T_c$.
}
\label{match_ma}
\end{figure}

We have performed a detailed analysis of 
propagators calculated in Landau gauge in 4d and 3d at $T=2 T_c$. The latter
has been simulated at values of $h$ corresponding to the metastable region
of the 3d adjoint Higgs model, {\it i.e.} for $\beta=11$, $h=-0.3773$ and
$x=0.099$ (see appendix for details).
The 4d simulations have been performed 
at $\beta=2.74$ on lattices given in Table I. 
In order to compare the propagators obtained from simulations in 
3d and 4d we have normalized them to unity at distance $z T=0$. The results are
shown in Figure \ref{match}. As one can see from the figure the 3d and 4d
results for the electric as well as for the magnetic propagators are in excellent
agreement.
It is interesting to note that good agreement between
3d and 4d propagators is found already at relatively short distances,
although the 3d effective theory is expected to describe the
4d physics only at distance $zT >1$.
The above result implies  that
dimensional reduction works quite well even at $T=2T_c$ and can be established
by comparing gauge fixed observables. 
In Figure \ref{match_ma} we compare the electric and magnetic propagators
calculated from full 4d and effective 3d theories at $2 T_c$ for the maximally
Abelian gauge. As one can see from the figure a good agreement between 4d and 3d
results exists also here, although it is a priori not clear in which sense
the 3d Maximally Abelian gauge corresponds to the 4d Maximally Abelian gauge. 
The agreement between the magnetic propagators in Figure \ref{match_ma} shows
that the two gauges are quite similar.  
A possible reason for this agreement is the decoupling of the $A_0$ field from the
dynamics of spatial (magnetic) gauge fields which will be discussed
in the next section.
We note, however, already here that the magnetic propagators calculated in
Landau
and Maximally Abelian gauges, respectively, show quite a different long distance
behavior. This is evident from the comparison of Figure 1 and Figure 2 and will
be analyzed in much more detail in the next section.
Also the apparent volume dependence visible in these figures will be 
discussed later in more detail.

\section{Numerical results on electric and magnetic gluon propagators}

In this section we will discuss our numerical results obtained from
simulations in three and four dimensions. The main purpose of this 
investigation was to quantify the gauge dependence of the electric
and magnetic gluon propagators and analyze to what extent gauge invariant
masses can be extracted from the long distance behavior of the gluon
correlation functions. 

\begin{figure}
\epsfxsize=6cm
\epsfysize=6cm
\centerline{\epsffile{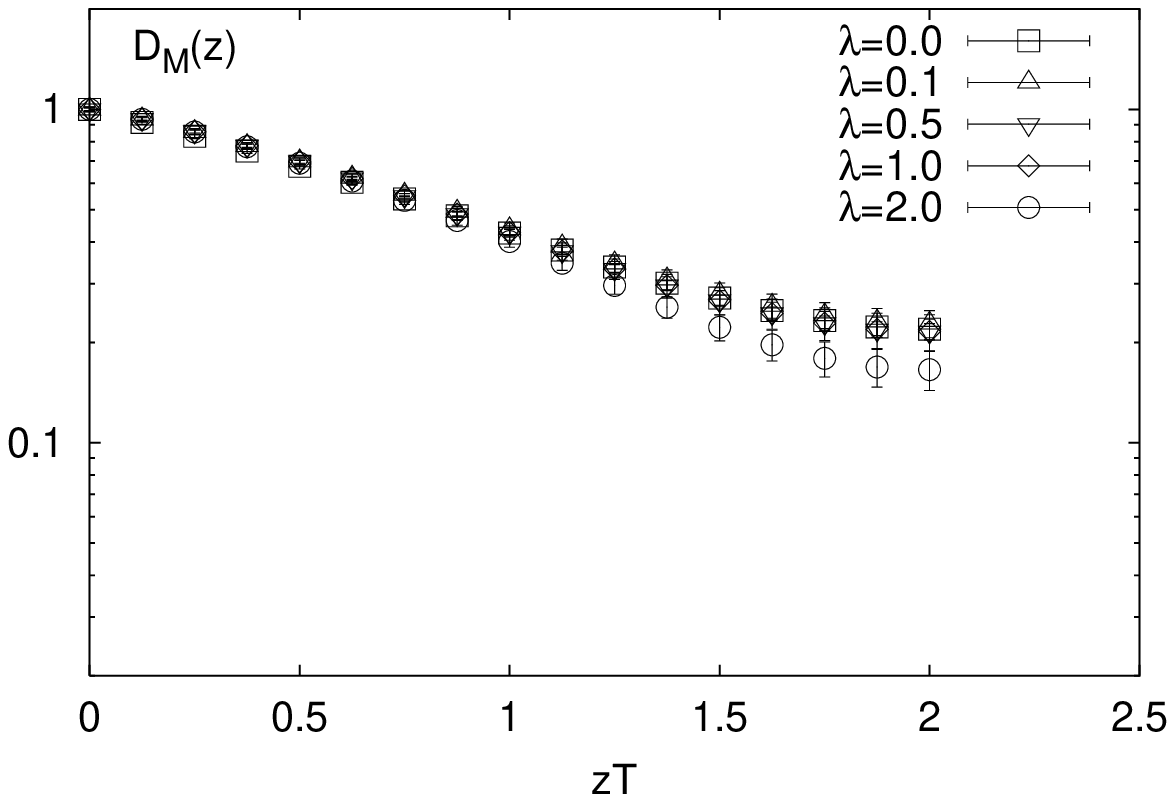}
\epsfxsize=6cm \epsfysize=6cm \epsffile{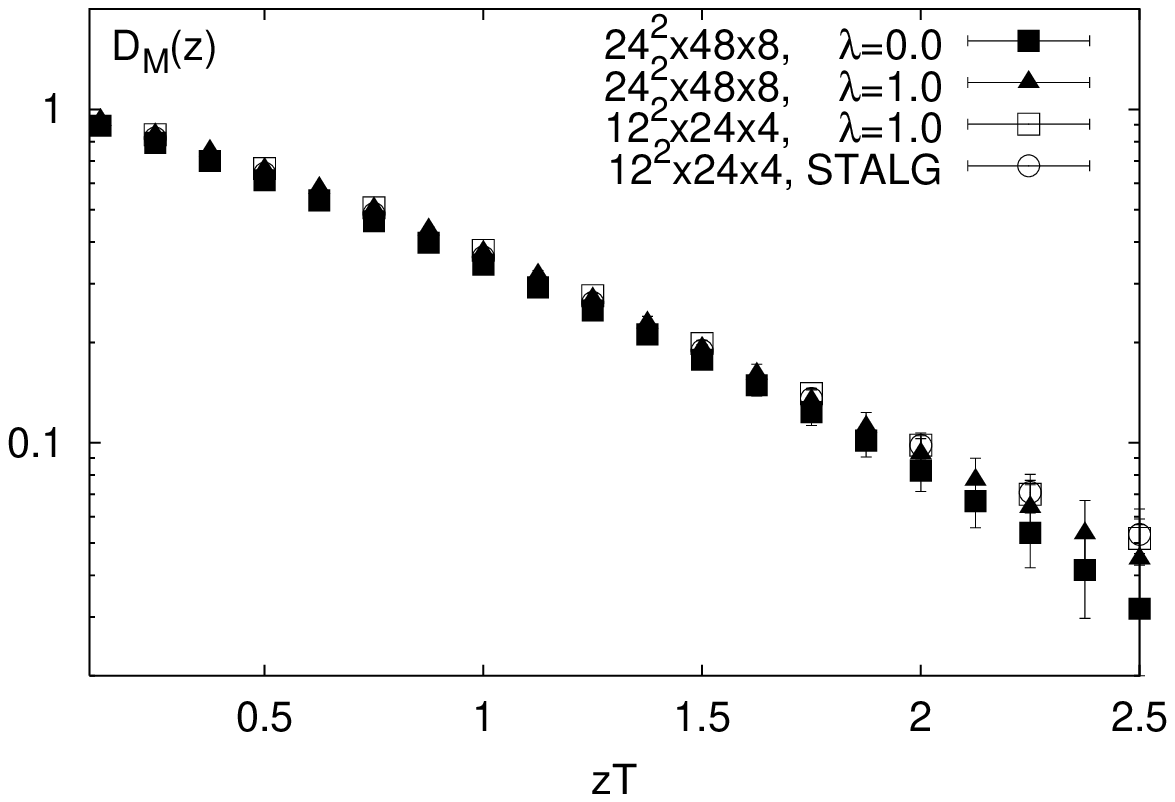} }
\caption{The magnetic propagator in $\lambda$ 
gauges on a $16^2 \times 32 \times 8$ lattice (left), as well as on
$24^2 \times 48 \times 8$ and $12^2\times 24 \times 4$  lattices (right) at $T=2T_c$.
Here also the propagator in STALG is shown.
The propagators were normalized to $1$ at $z=0$.  
}
\label{magn4d}
\end{figure}

In Figure \ref{magn4d} 
we show results from a calculation of the magnetic propagator
in 4d finite temperature gauge theory in different
$\lambda$-gauges including the Coulomb gauge limit.  
All the $\lambda$-gauges yield identical magnetic propagators. 
In fact, one should expect that all $\lambda$-gauges do lead to identical propagators
on an infinite spatial lattice. For static configurations ($\d_0 A_0=0$) the gauge 
condition for the $\lambda$-gauges, $\lambda \d_0 A_0+\d_i A_i=0$ is up to multiplicative 
factors identical
to the 3d Landau gauge condition. 
As we have shown that the 4d propagators in Landau
gauge can be mapped onto the 3d propagators this has to be the case also for the 4d
propagators calculated in $\lambda$-gauges. The same holds for STALG.

\begin{figure}
\epsfxsize=6cm
\epsfysize=6cm
\centerline{\epsffile{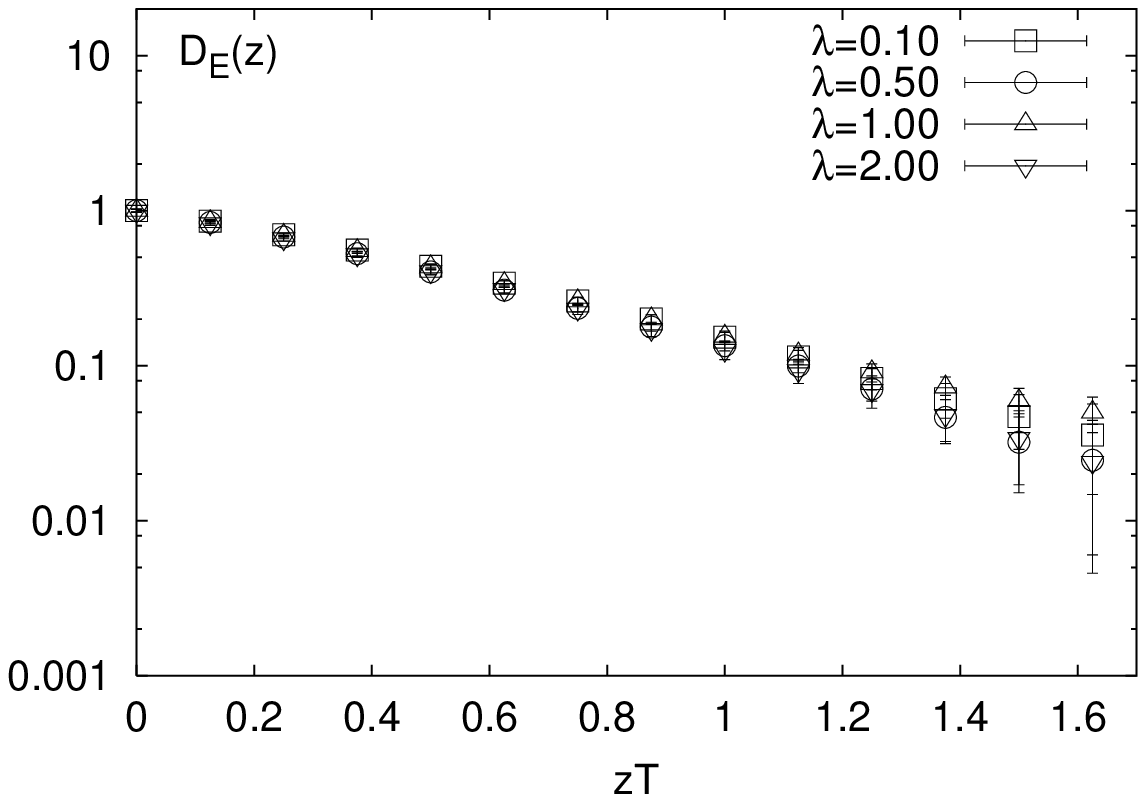}
\epsfxsize=6cm \epsfysize=6cm \epsffile{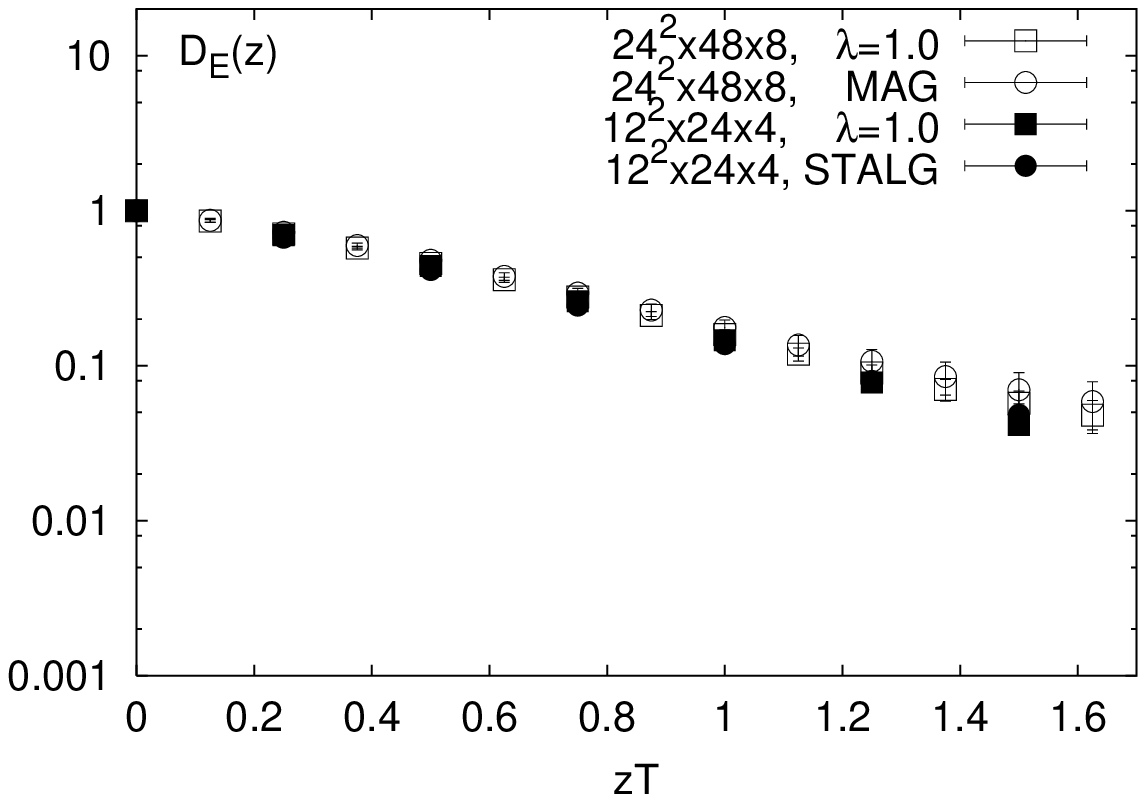}}
\caption{The electric propagators in different gauges.
Shown is the electric propagator measured on $16^2 \times 32 \times 8$ (left)
and on $24^2 \times 48 \times 8$ and $12^2 \times 24 \times 4$ (right). 
The propagators were normalized to $1$ at $z=0$.
}
\label{elec4d}
\end{figure}

We also note, that magnetic propagators show  strong volume dependence  
at distances $zT \gsim 1$. This effect may be traced back to the influence of zero mode
contributions to the propagators \cite{damm98} which are different in different gauges 
and give a volume dependent positive contribution to the correlation
functions. Zero mode
contributions are most prominent in the long distance behavior of the correlation 
functions calculated on finite lattices. 
In Figure \ref{elec4d} we summarize our results for electric propagators.
As in the case of magnetic propagators, electric propagators calculated in
different 4d $\lambda$-gauges as well as in STALG agree with each other
as expected.
Contrary to magnetic propagators the electric propagators show no significant 
volume dependence. We also note
that the cut-off effects seem to be small both for electric and magnetic
propagators (cf. Figures \ref{magn4d} and \ref{elec4d}).
The issues of volume and cut-off dependence will be discussed separately
for electric and magnetic propagators in the following two subsections.

\subsection{The magnetic gluon propagator}
In this section  we will discuss in more detail our results on magnetic propagators.
Previous lattice calculations of the magnetic propagators in hot SU(2) gauge theory in
4d \cite{heller98} and 3d \cite{karsch98} gave evidence for its exponential decay
in coordinate space and thus indicated the existence of a magnetic mass.  A non-zero 
magnetic mass was also found in analytical approaches based on gap equation 
\cite{alexanian95,buchmuller95,eberlein98} and a non-perturbative analysis of
$2+1$ dimensional 
gauge theory \cite{nair}. 
Nonetheless, it has been questioned whether a magnetic
mass in non-abelian gauge theories does exist \cite{jackiw96,cornwall98}. 
In Ref. \cite{cucchieri99} the Landau gauge propagators of 3d pure gauge theory
were studied on fairly large lattices and were found to be infrared suppressed
\footnote{Similar results were found in 4d SU(N) gauge theory at zero
temperature \cite{leinweber98,smekal98}.}.
Such a behavior clearly is in conflict with the existence of a simple pole mass.
In order to further
investigate this problem simulations on larger lattices 
are needed to explore the long distance regime of the correlation functions. 
This can be achieved in the 3d reduced theory which gives a good description of
the 4d theory even at temperatures a few times $T_c$. We thus concentrate, in the following
on a discussion of results 
obtained from our calculations in 3d. In fact, it also has been observed
already in earlier calculations that  
the magnetic propagators 
of the 3d adjoint Higgs model are very close to
the corresponding propagators of 3d pure gauge theory. 
We thus further restrict our analysis of the magnetic propagators to the limit
of 3d pure gauge theory. Where appropriate we will perform a comparison with
results obtained from
the 3d adjoint Higgs model and the 4d SU(2) gauge theory.

In order to get control over the propagators also in the continuum as well as in
the infinite volume limit we have performed calculations for
$\beta_3=5,~5.5,~6,~8$ and $16$ on different lattice volumes.
The simulation parameters are summarized in Table II.

\begin{center}
\vskip0.5truecm
\begin{tabular}{|c|c|c|c|c|}
\hline
$~\beta_3=5~$  &$~\beta_3=5.5~$   &$~~\beta_3=6~~~$ &$~~\beta_3=8~~~$ &$~~\beta_3=16~~$\\
\hline
$~10^2\times20~$ &$~~16^2 \times 32~~~$ &$~72^3$ &$~~16^2 \times 32~~~$ &$~~~32^2 \times 64~~$\\
$~~~16^3~~~$ & $~~24^2 \times 48~~~$ & $~~~~~$  &$~~24^2 \times 48~~~$  &$~~~40^2 \times 96~~$\\
$~~~24^3~~~$ & $~~$                  & $~~~~~$  & $~~28^2 \times 56~~~$ &$~~~48^2 \times 96~~$\\
$~~~30^3~~~$ & $~~$                  & $~~~~~$  & $~~32^3~~~$           &$~~64^3~~$\\
$~~~32^3~~~$ & $~~$                  & $~~~~~$  & $~~32^2 \times 64~~~$ &$~~96^3~~$\\
$~~~40^3~~~$ & $~~$                  & $~~~~~$  & $~~48^3$              &$~~$\\
$~~~48^3~~~$ & $~~$                  & $~~~~~$  & $~48^2 \times 64$     &$~~$\\
$~~~56^3~~~$ & $~~$                  & $~~~~~$  & $~64^3$               &$~~$\\
$~~~60^3~~~$ & $~~$                  & $~~~~~$  & $~96^3$               &$~~$\\
$~~~64^3~~~$ & $~~$                  & $~~~~~$  & $~~$                  &$~~$\\
$~~~72^3~~~$ & $~~$                  & $~~~~~$  & $~~$                  &$~~$\\
$~~~96^3~~~$ & $~~$                  & $~~~~~$  & $~~$                  &$~~$\\
\hline
\end{tabular}
\vskip0.5truecm
TABLE II: Lattice volumes used at different values of $\beta_3$ in our 3d 
calculations of the magnetic gluon propagator.
\vskip0.5truecm
\end{center}

It has been pointed out in \cite{damm98} that zero modes can give sizeable contributions
to gluon correlation functions. In terms of the lattice gauge fields the zero mode 
contribution is defined as the expectation value of the average gauge field, 
\be
\phi_{\mu}^a  = {1\over \Omega} \sum_x A_{\mu}^a(x) \quad .
\ee
The zero mode contribution is apparent in the momentum space propagators where one
finds from eq.~\ref{fmu} for vanishing momentum 
\be
\tilde{D}_{\mu \mu}(p=0)=a^d \Omega \sum_a \langle  {(\phi_{\mu}^a)}^2 \rangle \quad .
\ee
 
\begin{figure}
\epsfxsize=8cm
\epsfysize=7cm
\centerline{\epsffile{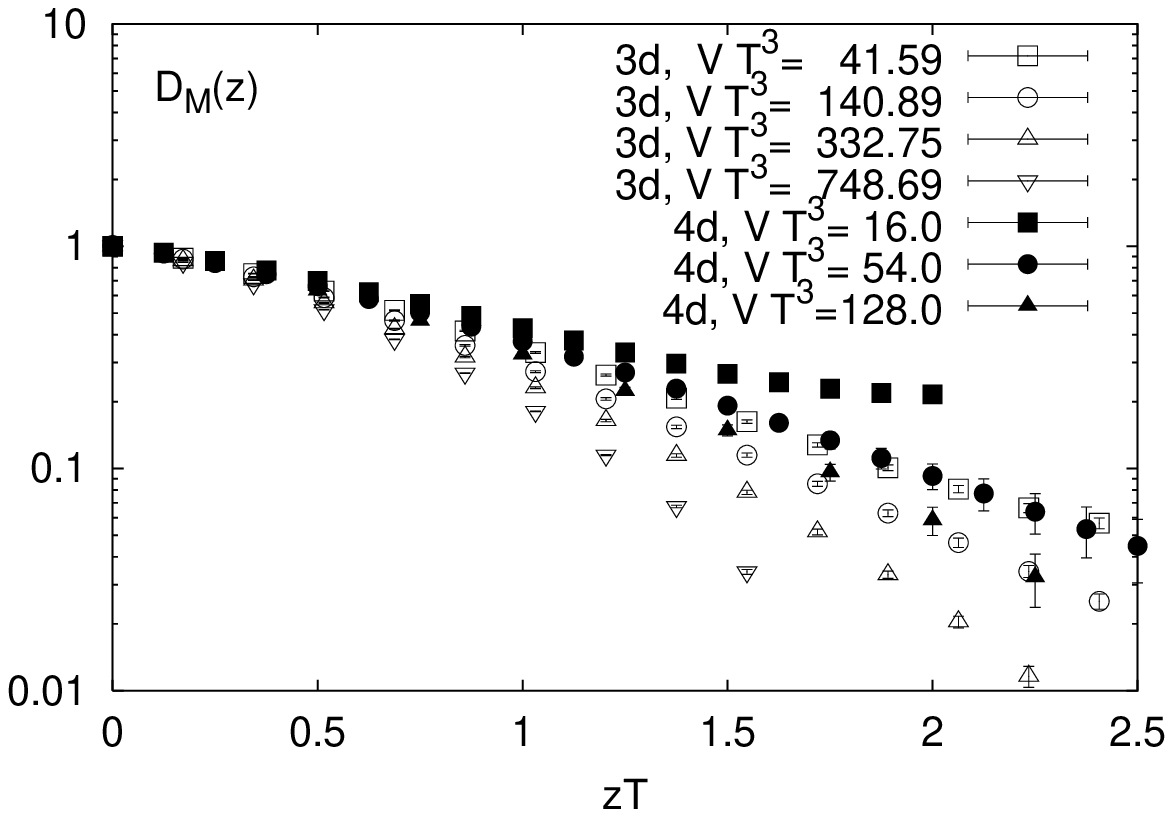} \epsfxsize=8cm \epsfysize=7cm
\epsffile{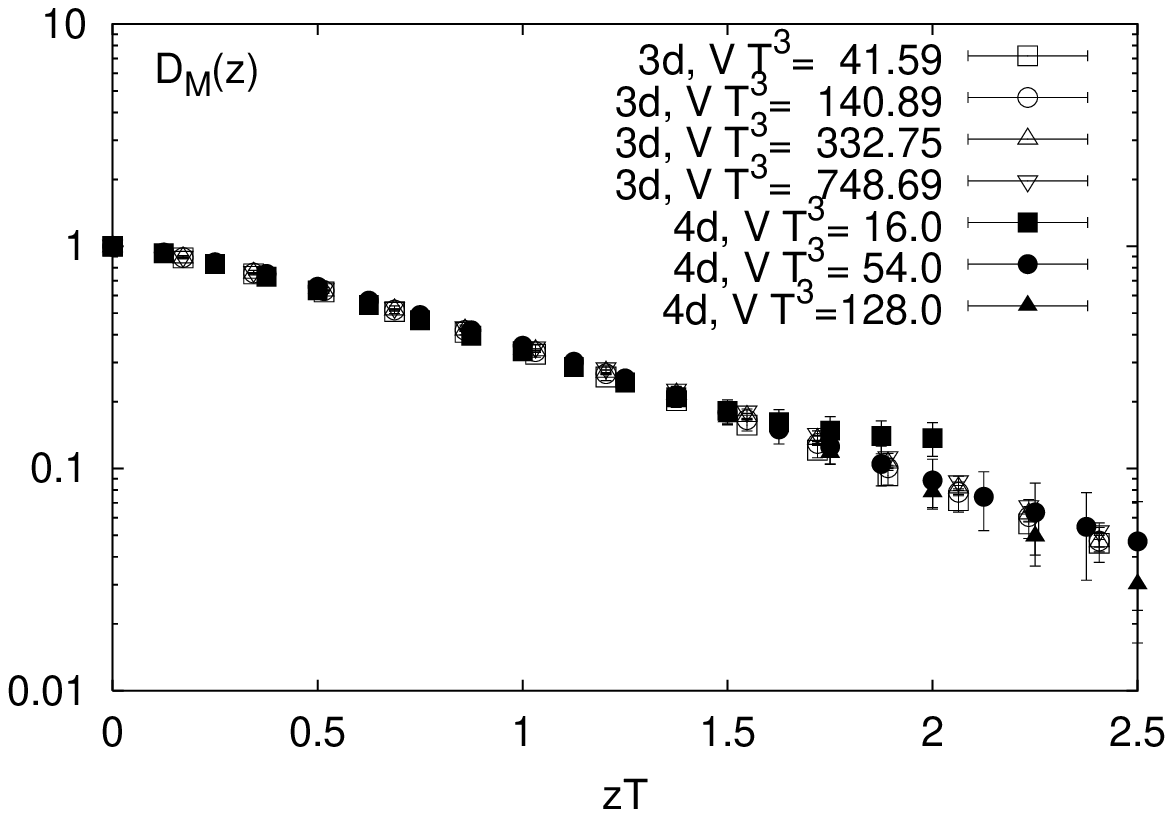}}
\caption{
Volume dependence of the magnetic propagator in coordinate 
space calculated in Landau (left) and Maximally Abelian (right) gauges.
The propagators were normalized to $1$ at $z=0$. Shown are 
results from 3d simulations at $\beta_3=8$ 
and 4d simulations, which are compared at similar 
values of the physical volume in units of $T^3$. These volumes correspond to
lattices of size $16^2\times 32 \times 8$, $24^2 \times 48 \times 8$ and 
$16^2 \times 32 \times 4$ in 4d and to $16^2 \times 32$, $24^2 \times 48$, 
$32^2 \times 64$ and $48^2 \times 64$ in 3d. The $\beta_4$ values for our
4d simulations are given in Table I.
} 
\label{magn_n_dep}
\end{figure}

By taking the inverse Fourier transformation of Eq.(\ref{fmu}) one can easily
see that the zero modes give a constant contribution ${(N_z a)}^{-1} \tilde{D}_{\mu \mu}(p=0)$
to the coordinate space propagator $D_{\mu \mu}(z)$.
In fact, 
such a contribution does qualitatively explain the overall volume dependence
of the magnetic propagators observed in our calculation.
The contribution from zero momentum mode fluctuations is expected to vanish in the 
infinite volume limit.  We thus expect that at fixed distance
$zT$ the propagators approach their asymptotic, infinite volume values
from above as the lattice size is increased. This is indeed the case, as can be
seen in Figure \ref{magn_n_dep} where we show the 4d and 3d  magnetic
propagators calculated in Landau and Maximally Abelian gauges on different size 
lattices at $T=2T_c$. As one can see from the figure the volume dependence is
quite different in these two gauges. 
In the case of Landau gauge we observe that the 
magnetic propagators in coordinate space decay faster with increasing lattice size.
In the Maximally Abelian gauge, however, the propagator does not exhibit
any sizeable finite size dependence.
\vspace*{-0.3cm}
\begin{figure}
\epsfxsize=8cm
\epsfysize=7cm
\centerline{\epsffile{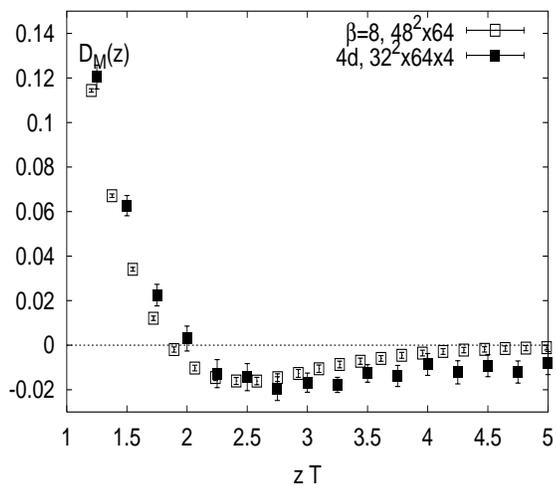}}
\caption{Large distance behaviour of the magnetic propagator in coordinate space
calculated in Landau gauge.
The propagators were normalized to 1 at $z=0$. 
} 
\label{demo}
\end{figure}
For volumes $V T^3 \gsim 300$ the magnetic propagator calculated in Landau
gauge becomes negative
for $z T \gsim 2$. In Figure \ref{demo} we show the large distance behavior of
the coordinate space
propagator calculated in 4d and 3d on lattices with similar
3d lattice volume. A similar behavior of the coordinate space propagator
was found in other $\lambda$-gauges.
The propagators calculated in MAG, however, stay positive for all lattice sizes
and distances accessible in our calculation, i.e. up to $z T \sim 5$.

\begin{figure}
\epsfxsize=6.5cm
\epsfysize=5.4cm
\centerline{\epsffile{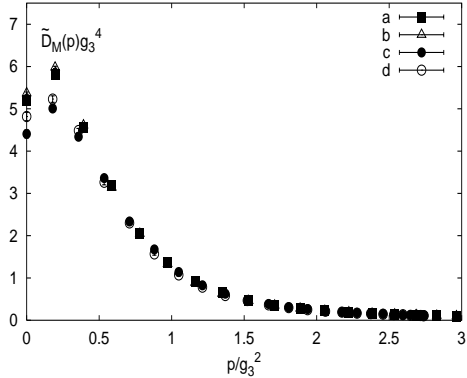}
}
\caption{
Momentum space magnetic propagator obtained from 3d pure gauge
theory and the adjoint Higgs model with couplings corresponding
to temperatures $2 T_c$ and $3 T_c$. Shown are
the magnetic propagators in pure gauge theory at $\beta_3=8$ (a),
in 3d adjoint Higgs model at $\beta_3=8$, $x=0.09,~h=-0.4846$ corresponding
to the temperature $3 T_c$ (b), 
in pure gauge theory at $\beta=5.5$ (c)
and in 3d adjoint Higgs model at $\beta=5.5$, $x=0.099,~h=-0.7528$ corresponding
to the $2 T_c$ (d).
}
\label{p_adj}
\end{figure}

As was mentioned above the magnetic
propagators in the  3d adjoint Higgs model are not very sensitive to the presence of
the adjoint Higgs field. This is illustrated in Figure \ref{p_adj} where we
compare magnetic propagators in momentum space calculated in the 3d adjoint
Higgs model with corresponding results obtained in the pure gauge theory at the 
same value of $\beta_3$.
This also shows that the effect of the adjoint
Higgs field on the magnetic propagator decreases with increasing temperature.
In fact, for temperature $T \gsim 10 T_c$ we find 
no visible effect of the adjoint Higgs field on the 
magnetic propagators. 
In what follows we will therefore mainly discuss the magnetic propagators in
the limit of pure gauge theory.

\begin{figure}
\epsfxsize=7.5cm
\epsfysize=6.3cm
\centerline{\epsffile{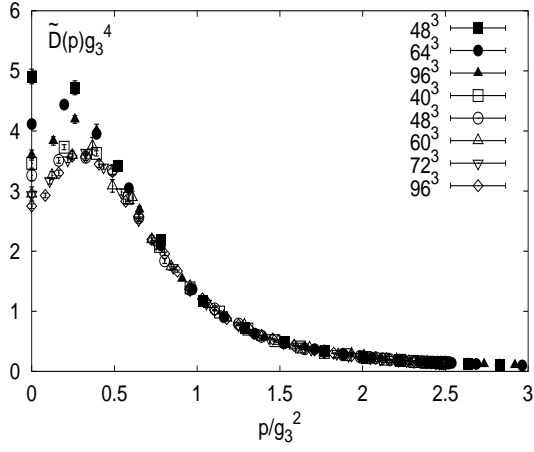}
}
\caption{
The momentum space magnetic propagator at $\beta_3=5$ (open symbols) 
and $\beta_3=8$ (filled symbols) calculated in Landau gauge. 
} 
\label{magnp3d}
\end{figure}

In Figure \ref{magnp3d} we show the momentum space magnetic propagators in Landau gauge 
obtained from 3d pure
gauge theory at $\beta_3=5$ and $\beta_3=8$. 
For momenta $p>0.6g_3^2$ the propagators in units of $g_3^4$ 
are volume and $\beta_3$-independent. For small momenta, however,
they are strongly volume  and also $\beta_3$ dependent.
One can clearly see that for large volumes the propagators  reach a
maximum at non-zero momentum and start decreasing with decreasing
momenta. This is a direct consequence of the negative propagators in coordinate
space found in Landau gauge for $z T \gsim 2$.
 
\begin{figure}
\epsfxsize=7.5cm
\epsfysize=6.3cm
\centerline{\epsffile{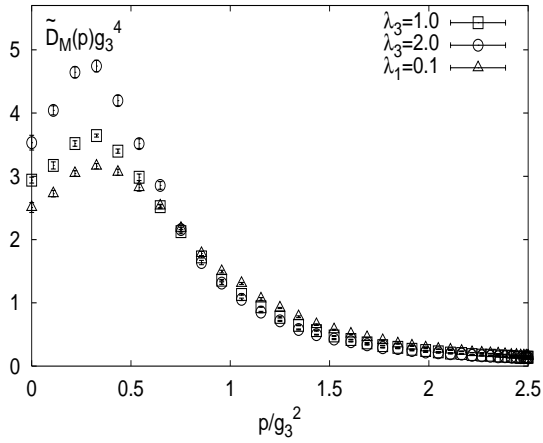}
}
\caption{
The momentum space magnetic propagator at $\beta_3=5$ on $72^3$ lattice 
in different $\lambda$-gauges.
} 
\label{magnp3d_b5g}
\end{figure}

The volume dependence of the magnetic propagator in momentum space
is strongest at $p=0$. On the other hand at $\beta_3=5$ one can see that the value of the
propagator in the vicinity of its maximum is essentially volume independent for the 
three largest volumes. A similar behavior of the momentum space
propagator was found in other $\lambda$-gauges. This is demonstrated by Figure
\ref{magnp3d_b5g}. Note that although the propagator is a gauge dependent
quantity the position of the peak seems to be gauge independent
in the class of $\lambda$-gauges considered here.

\begin{figure}
\epsfxsize=7.5cm
\epsfysize=6.3cm
\centerline{\epsffile{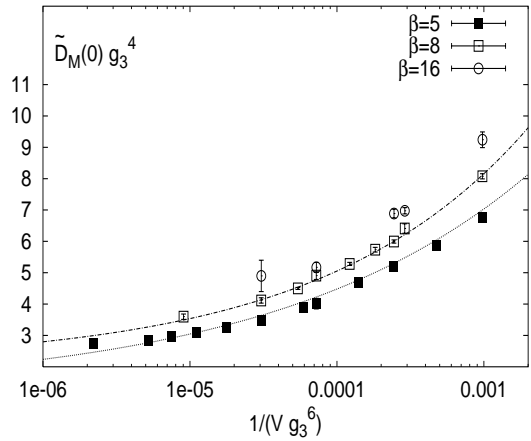}
}
\caption{
Volume dependence of $D_M(0)$ at different values of $\beta_3$.
The lines represent  fits with the ansatz given in Eq.(\ref{pscale}). 
} 
\label{zero3d}
\end{figure}

The existing rigorous bound on the infrared behavior of the 
gluon propagator in momentum space
imply that it is less singular than $p^{-2}$ in 4d and $p^{-1}$ in 
3d \cite{zwanziger91}. It has been further argued that it is likely
to vanish at zero momentum in the thermodynamic limit \cite{zwanziger94}.
These considerations  were also extended  to the case of magnetic propagators in gauge
theories at finite temperature \cite{zahed00}. 
We have analyzed this question by performing studies of the volume dependence of
the zero momentum propagator.
The data for $\tilde{D}_M (0)$ are shown in Figure \ref{zero3d}
for different $\beta_3$ values. For $\beta_3=5$ and $\beta_3=8$ the data
have been fitted to the ansatz
\be
\tilde{D}_M(0) g_3^4 =  a (V g_3^6)^{-z} + b \quad .
\label{pscale}
\ee
We have performed fits with $b=0$ and $b>0$.
In the first case we have obtained rather large $\chi^2$ values 
(typically the value of $\chi^2/d.o.f$ was between 4 and 10)
for both $\beta_3=5$ and $\beta_3=8$.
The three parameter fits give a reasonably good $\chi^2$ for both $\beta_3$ values.
Using the whole range of volumes
one finds $a=51(11), b=2.06(15)$ and $z=0.34(3)$ with $\chi^2/d.o.f.=1.8$ 
for $\beta_3=5$ and  $a=52(9),b=2.08(33)$ and $z=0.31(3)$ with with
$\chi^2/d.o.f.=1.8$ for  $\beta_3=8$.
Although the volume dependence of $\tilde{D}_M(0)$ seems to be
$\beta_3$-independent, the value of the exponent $z$ as well as the
value of $\tilde{D}_M(0)$ in the infinite volume limit strongly
depends on the range of volumes used in the fit. For example fitting the
data on $\tilde{D}_M(0)$ for $\beta_3=5$ using lattice volumes from $30^3$ 
to $96^3$ one gets : $a=281(139),b=2.52(56)$ and $z=0.55(5)$. 
In order to firmly determine the functional form of the volume
dependence of $\tilde{D}_M(0)$ and the value $\tilde{D}_M(0)$ in the
infinite volume limit simulation on even larger lattices are necessary.

\begin{figure}
\epsfxsize=7.5cm
\epsfysize=6.5cm
\centerline{\epsffile{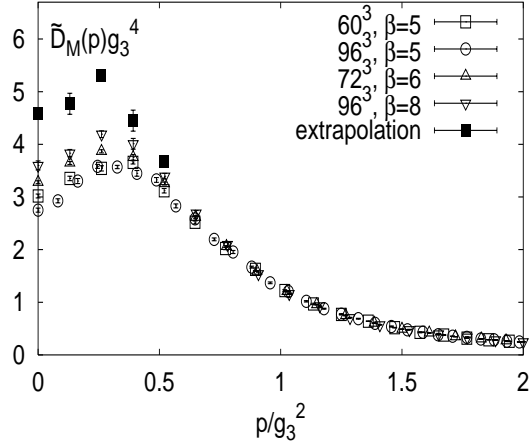}
}
\caption{The Landau magnetic propagator at different values of $\beta_3$ and
the continuum extrapolation.
} 
\label{magnp3d_cont}
\end{figure}

In order to confirm the infrared
suppression of $\tilde{D}_M(p)$ and the existence of a maximum in
$\tilde{D}_M(p)$
at non-zero momentum we have analyzed in more detail the interplay between
infinite volume and continuum limit. 
\vspace*{-0.2cm}
We have seen that the propagators are infrared
suppressed for large enough volumes at $\beta_3=5$ and $8$. However, the magnitude of
the propagator at small momenta shows a clear $\beta_3$-dependence. Therefore,
it is important to estimate the behavior of the magnetic propagator in 
the continuum limit. We have seen that at $\beta_3=5$ the magnetic propagator on
a $60^3$ lattice is already close enough to the corresponding infinite
volume limit. Therefore we have performed additional calculations of the magnetic 
propagators at $\beta_3=6$ and on $72^3$. The lattice volumes $60^3$, $72^3$ and
$96^3$ at $\beta_3=5,~6$ and $8$ correspond to the same physical volume.
Moreover, for small momenta $p<0.5 g_3^2$, values of the propagators at
approximately same values of momenta are available. In this region of momenta 
we have performed an extrapolation to the 
continuum limit fitting data at different $\beta_3$ with the ansatz 
$a+b/\beta_3$. The result of this analysis is summarized in Figure \ref{magnp3d_cont}.
As one can see from this figure the general structure of the propagator, in
particular the existence of a maximum at non-zero momentum 
is preserved in the continuum limit, although the volume dependence of the
momentum space propagator rapidly increases with decreasing momentum for
$p/g_3^2 \le 0.5$.

\subsection{The electric gluon propagator}

Let us now turn to a discussion of the electric gluon propagator.
The volume dependence of the electric propagator turns out to be quite different from
the magnetic ones.
The electric propagators show exponential decay at large distances in all
gauges considered.

\begin{figure}
\epsfxsize=7cm
\epsfysize=6cm
\centerline{\epsffile{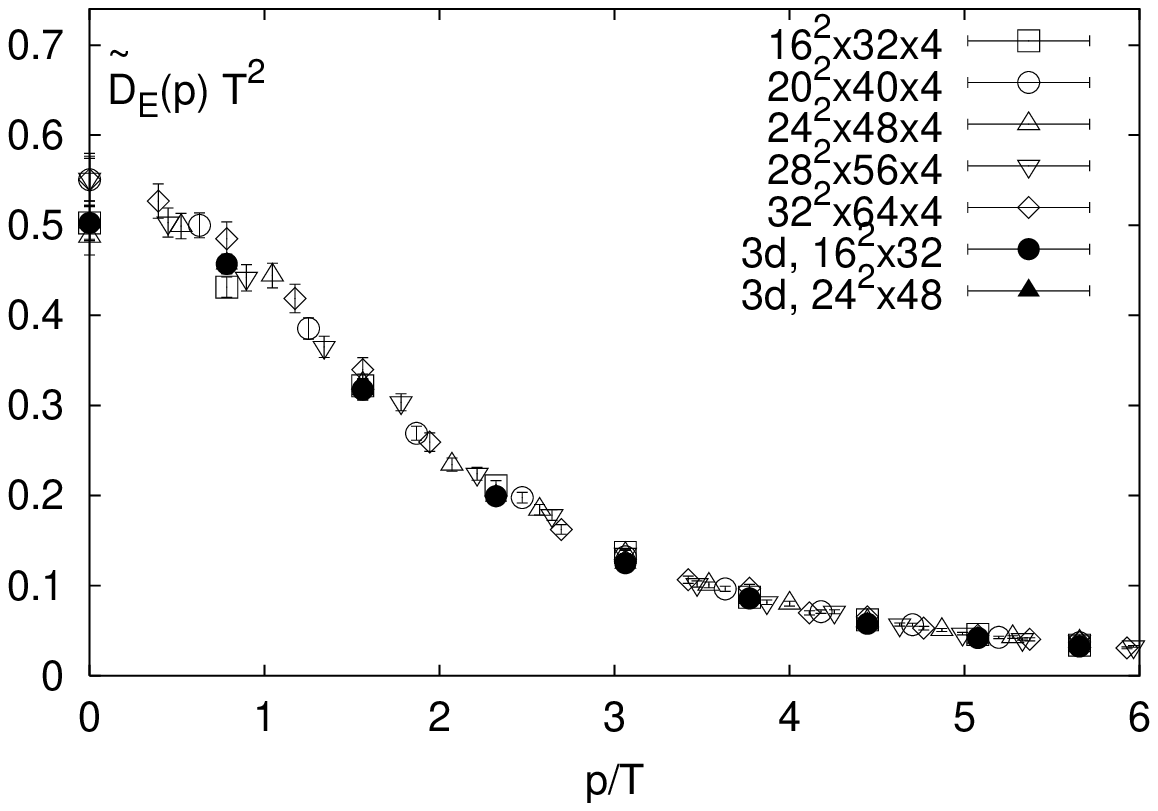}
\epsfxsize=7cm \epsfysize=6cm \epsffile{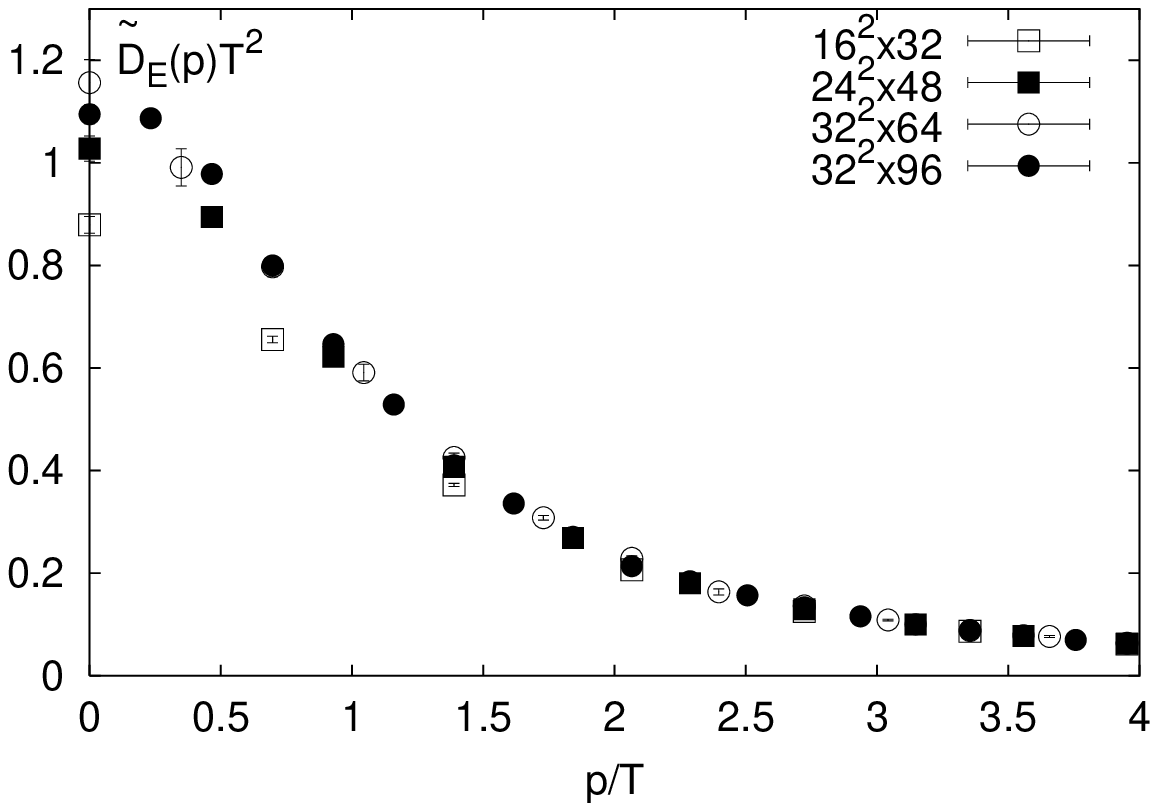}}
\caption{
The Landau gauge electric propagator in  momentum space at $2 T_c$ (left)
and $T=9200 T_c$ (right). 
}
\label{elecp}
\end{figure}

In Figure \ref{elecp} the Landau gauge electric propagators in momentum space
are shown for $T=2 T_c$ and $T=9200 T_c$. In the latter case we performed the analysis
only in the 3d effective theory at $x=0.03$, $\beta=16$ and $h=-0.2085$. 
As one can see from the figure the volume dependence of the electric propagators
is indeed quite different from the volume dependence of the magnetic propagators.
In particular the electric propagators show no sign of infrared suppression.
For fixed lattice geometry, i.e.  fixed ratio $N/N_z$
the volume dependence of the zero mode contribution seems to approach  $V^{-1}$
as expected for the propagator of a massive particle.  Moreover, the infinite 
volume limit is almost reached on the largest lattices used in our simulations.
We find a similar  volume dependence also in other gauges. 

\begin{figure}
\epsfxsize=7cm
\epsfysize=6cm
\centerline{\epsffile{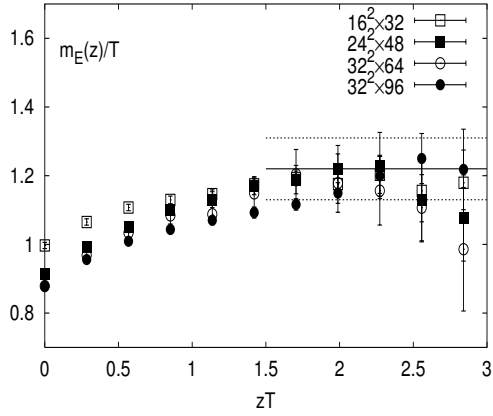}}
\caption{
Local electric masses in Landau gauge for $\beta=16$, $h=-0.2085$
and $x=0.03$. The solid line is  the value of the electric 
mass obtained from an exponential 
fit of the electric propagator calculated on a $32^2\times 96$ lattice.
The dashed lines indicate the uncertainty in its value.
}
\label{localme_fss}
\end{figure}

Since we are interested in extracting the screening mass from
the electric propagator it is also important to address the question of 
volume dependence of the electric mass. 
In Figure \ref{localme_fss} the local electric
masses in Landau gauge are shown for different lattice volumes 
for $\beta_3=16$, $h=-0.2085$ and $x=0.03$.
As one can see from the figure,
the value of the plateau (which in fact determines the screening mass) is essentially
volume independent, although there is some volume dependence in the 
local masses at short distances.
Thus the screening masses can be estimated even from
the smallest lattice for these values of the parameters.
We will use this fact in the discussion of the gauge dependence of the 
electric screening mass.

Since the electric propagators show exponential decay at large
distances in all gauges considered it is natural to ask 
whether the electric masses extracted from them are gauge
independent.
There is a formal proof that poles of finite temperature 
propagators are gauge independent to any order of perturbation theory
\cite{kobes90}. However, as  discussed in section I the Debye mass
is not calculable in perturbation theory beyond leading order.
Therefore it is not clear whether the arguments of Ref. \cite{kobes90} apply
for electric screening mass. In terms of the effective 3d theory the screening
of static electric fields is related to propagation of the adjoint scalar field.
Since the 3d theory is confining the adjoint scalar field is not a
physical state ( the physical states are some bound states) and therefore
there is no physical principle which guarantees its mass to be gauge invariant. 
Guided by the analogous problem in 3d scalar QED it was
conjectured in Ref. \cite{blaizot96} 
that the large distance behavior of the high temperature
non-Abelian electric propagator is
dominated by a brunch cut singularity. 
Although this is different from the simple pole obtained in leading order
perturbation theory this still leads to an exponential decay of the electric
propagators in coordinate space with
a gauge invariant screening mass. 
Thus it is clear that also the question of gauge (in)dependence of
the electric screening mass is non-trivial in general.  

We have studied the gauge dependence of the electric propagators for
two sets of parameters of the 3d effective theory 
$x=0.03~,h=-0.2085,~\beta_3=16$ and $x=0.03,~h=-0.1510,~\beta_3=24$ both
corresponding to the temperature $T=9200 T_c$. In this investigation lattices
with spatial volume $16^2 \times 32$, $24^2 \times 48$ and $32^2 \times 96$
were used. From studies of the Landau gauge propagators we know that electric
masses for $\beta_3=16$ can be reliably extracted already from $16^2 \times 32$
lattice.
We have investigated 
electric propagators in different gauges introduced in section II.B. For
$\beta_3=16$ most simulation were performed on $16^2 \times 32$ lattice. For 
calculations with $\lambda_3$
gauges we also used a $32^2 \times 96$ lattice. 
For $\beta_3=24$ propagators
in all gauges except the Maximally Abelian gauge were measured on
$32^2 \times 96$ lattice. Due to the  
large number of iterations required for fixing the Maximally Abelian gauge 
propagators were  measured
only on a $24^2 \times 48$ lattice  at $\beta_3=24$.
\begin{figure}
\epsfxsize=7cm
\epsfysize=6cm
\centerline{\epsffile{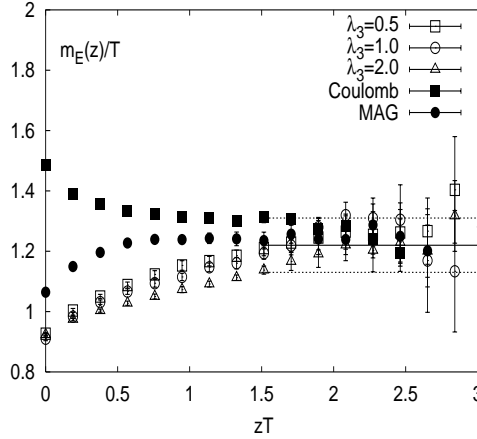}}
\caption{
The local electric masses calculated on $32^2 \times 96$ 
for $x=0.03$, $h=-0.1510$ and
$\beta_3=24$ in $\lambda_3$-gauges,
in 3d Coulomb gauge ($c_1=1$) and in Maximally Abelian gauge.
The solid line is  the value of the electric
mass obtained from an exponential
fit of the electric propagator in Landau gauge calculated on a $32^2\times 96$ lattice
for $\beta=16$, $h=-0.2085$
and $x=0.03$. 
The dashed lines indicate the uncertainty in its value.
}
\label{locmab24}
\end{figure}

Our main results for $\beta_3=24$ are summarized in Figure \ref{locmab24}, 
where local electric
screening masses in Coulomb, Maximally Abelian and $\lambda_3$- gauges
are shown. As one can see from the figure 
local masses are strongly gauge dependent at short distances. For $z T\gsim 2$
they, however approach a plateau which is gauge independent within the
statistical accuracy reached in our calculation.

\begin{figure}
\epsfxsize=7cm
\epsfysize=6cm
\centerline{\epsffile{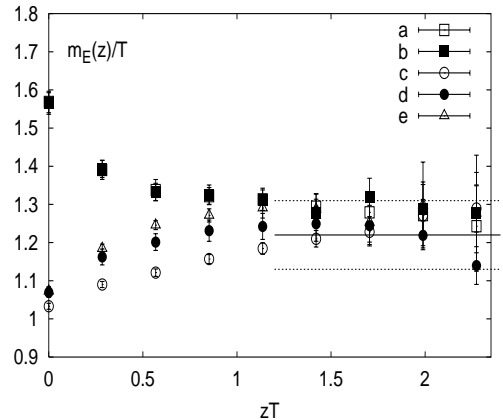}}
\caption{
The local electric masses calculated for $x=0.03$, $h=-0.2085$ and
$\beta_3=16$ on $16^2 \times 32$ lattice. 
Shown are the local electric masses calculated in $c_1$
gauges for $c_1=0.1$ (a) and $c_1=10$ (b), in $\lambda_1$ gauges
for $\lambda_1=0.1$ (c) and $\lambda_1=10$ (d) and Maximally Abelian gauge
(e).
The solid line is  the value of the electric
mass obtained from an exponential
fit of the electric propagator in Landau gauge calculated on a $32^2\times 96$ lattice
$\beta=16$, $h=-0.2085$
and $x=0.03$.
The dashed lines indicate the uncertainty in its value.
}
\label{locmab16}
\end{figure}

We also have  performed simulations on a $16^2 \times 32$ lattice
for $\beta_3=16$ using Coulomb gauges with $c_1=0.1,~10$, as well as 
$\lambda_1$ -gauges with $\lambda_1=0.1,~10$
and Maximally Abelian gauge. The corresponding results for local electric
masses are shown in Figure \ref{locmab16}. 
Again within statistical accuracy no
gauge dependence of the plateau of the local electric masses can be 
observed, though the local masses have quite different $z$-dependence.

\section{Conclusions}

We have investigated  electric and magnetic gluon propagators in the
high temperature phase of SU(2) pure gauge theory in different gauges. 
The propagators
were calculated directly in full 4d theory and also in 3d effective theory.
It was shown that the effective theory can describe the electric and
magnetic propagators remarkably well down to temperature $T=2 T_c$.
We find that electric propagator can safely be extrapolated to the infinite
volume limit. 
within the statistical accuracy of our present analysis and the class of gauges
considered here its long distance behaviour is gauge independent and yields a
gauge independent electric screening mass which is compatible with
earlier determinations.

The magnetic propagator, however, has a complicated volume and gauge dependent
infrared structure which is not compatible with a simple pole mass.
While it still shows a simple exponential decay at large distances in 
the Maximally Abelian gauge it starts getting negative for $z T \gsim 2$ in a 
class of $\lambda$-gauges which includes the Landau gauge. This leads
to the infrared suppression of the propagator in momentum space and is 
incompatible with a simple pole in the magnetic gluon propagator as it
was deduced from earlier studies \cite{karsch98,heller95,heller98,karsch96}   
which were limited to shorter distances and smaller lattices.
Nonetheless, we find in all gauges that magnetic correlators are screened at
large distances.

\vspace{0.5cm}
\noindent
{\bf Acknowledgements:}

\medskip
\noindent
This work has been supported by the TMR network Finite Temperature
Phase Transition in Particle Physics (EU contract no.
ERBFMRX-CT-970122) and by the DFG under grant Ka 1198/4-1.
Our calculations have been partially performed at the HLRS in
Stuttgart and the $(PC)^2$ in Paderborn. The work of A.C.
also has been supported by FAPESP (Proj. no. 00/05047-5).
\section*{Appendix}

We will discuss here the procedure we followed to fix the couplings of the
3d adjoint SU(2) Higgs models.
Two different approaches were proposed for fixing the parameters
appearing in (\ref{act}), the usual perturbative dimensional reduction
\cite{kajantie97a,lacock92} and non-perturbative matching analyzed in
\cite{karsch98}. Here we have used both approaches.

The lattice gauge coupling $\beta_3$ is related to the
dimensionful 3d continuum gauge coupling $g_3^2$ by the standard
relation
\be
\beta_3={4\over g_3^2 a}.
\ee
This relation basically determines the lattice spacing $a$ in terms of the
dimensionful coupling $g_3^2$. The Higgs self-coupling and
the 3d gauge couplings are related to the renormalized gauge coupling
$g(\mu)$ of SU(2) gauge theory in $\overline{MS}$ scheme. The corresponding
relations were calculated in Ref. \cite{kajantie97a} to 2-loop level.  

To fix the temperature scale one needs to choose the renormalization
scale $\mu$ which fixes the temperature dependence 
of the 4d gauge coupling. For this we will use information on the
temperature dependence of the spatial string tension. 
The analysis of Ref. \cite{caselle94} showed that the spatial string
tension can be well described in terms of 3d effective theory.
Furthermore it was observed
that the temperature dependence of the 
spatial string tension of finite temperature
SU(2) gauge theory can well be described by a simple formula
$0.334(14)g^2(T)T$ \cite{bali93} with $g(T)$ given by 
the 1-loop renormalization group relation. 
The string tension of the pure 3d SU(2) gauge theory was found to
be $0.335(2)g_3^2$ \cite{teper99}. 
The string tension of the 3d adjoint SU(2) theory
has been calculated in \cite{hart99}. Within statistical errors it turned out 
to be independent of the scalar couplings and equal to the string tension of 
the pure SU(2) gauge theory. Based on these observations we have
fitted the data on the spatial string tension from \cite{bali93} using
$0.335(2)g_3^2$ with 1-loop level coupling $g_3^2$ and the relation
$T_c=1.06 \Lambda_{\overline{MS}}$ from \cite{heller98}.
The fit yields $\mu=18.86 T$. 
We have also fitted the data for the spatial string tension 
using the 2-loop formula for $g_3^2$ from \cite{kajantie97a}.
However, it turned out that the simple 1-loop formula fits the 4d
data on spatial string tension much better. Based on these observation 
the 3d gauge coupling $g_3^2$ and the Higgs self-coupling were chosen
according to the 1-loop version of formulas (2.13) and (2.15) from
\cite{kajantie97a}
\ba
&&
g_3^2=g^2\left(\mu\right) T,\\ 
&&
x={g^2\left(\mu\right)\over 3 \pi^2},
\label{g32x}
\ea
with $\mu=18.86 T$. 
In Ref. \cite{karsch98} the parameters $g_3^2$ and $x$ were chosen
according 2-loop formulas from \cite{kajantie97a} and $\mu=2 \pi T$.
We have checked that differences between the present choice and 
that used in  \cite{karsch98} are about $10 \%$.
In Ref. \cite{hart99} the parameters of the effective theory
were also calculated using the 2-loop formulas from \cite{kajantie97a},
however, the 4d gauge coupling constant was determined from
1-loop formula  in $\overline{MS}$-scheme
with $\mu \sim 7.0555$.
In Ref. \cite{hart99}
the screening masses extracted from gauge invariant correlators
were calculated in the effective 3d theory and compared with the
corresponding results from 4d simulations \cite{datta98,datta99}.
A rather good agreement between the screening
masses calculated in 3d effective and full 4d theories was found. 
However, the  3d gauge coupling constants in \cite{hart99}
were considerably larger than ours which
would overestimate the spatial string tension by  $30\%$.
Nevertheless this contradiction can easily be resolved by noticing
that the set of parameters corresponding to $5 T_c$ in \cite{hart99},
namely $x=0.104$ and $y=0.242$ would roughly correspond to the temperature
$2 T_c$ in our approach. Using the results from \cite{hart99} for these
values of $x$ and $y$ and
the value of $g_3^2(2 T_c)=2.89T$ we get for the two smallest
screening mass the values $m(0^{+}_{+})=2.90(3)$ and $m(0^{-}_{-})=3.91(8)$
(for details about classification of different gauge invariant screening masses
see Ref. \cite{hart99}). These numbers should be compared with those obtained
in 4d simulation in \cite{datta99} at $T=2 T_c$,  $m(0^{+}_{+})=3.06(12)$ and
$m(0^{-}_{-})=4.06(12)$. Thus it seems that with our present choice
of the parameters the effective 3d theory
could describe all static quantities measured so far in lattice simulations.
We also note that in  the case of 2+1-dimensional gauge theory at finite temperature
it has been shown that  the spatial string tension and
gauge invariant screening masses can be simultaneously 
well described by the effective 2d theory \cite{bialas00}. In 2+1
dimensions the situation is simplified by the fact, that the
gauge coupling has no renormalization scale dependence due to
superrenormalizability of the theory. 

Generally we have used the non-perturbatively determined
values for $h$ \cite{karsch98}. However, in some cases the propagators 
were also calculated for values of $h$ given by perturbative 
dimensional reduction. The bare Higgs mass parameter $h$ could
be related to the mass parameter $y=m_{D0}^2/g_3^4$ of the
continuum adjoint Higgs model \cite{kajantie97a}  
\be
h={16\over \beta_3^2}y-{3.1759114(4+5 x)\over \pi \beta_3}-
{1\over \pi^2 \beta_3^2}
\biggl( (20x-10x^2) (\ln{3\over 2} \beta_3+0.09)+8.7+11.6 x \biggr),\label{h4d} 
\label{h2y}
\ee
with $m_{D0}$ being the tree-level (from the point of view of the effective
theory) Debye mass. 
At 1-loop level one has:
\be
y={m_{D0}^2\over g_3^2}={2 \over 9 \pi^2 x}.
\label{y1}
\ee
From Eqs. (\ref{h2y}) and (\ref{y1}) we have calculated values of
$h$ corresponding to the 1-loop level dimensional reduction.
In this case simulation were performed in the
metastable brunch of the broken phase (see Ref. \cite{kajantie97a}
for details).

Numerical values for the couplings of the 3d effective action used in most of
our calculations are summarized in 
Table III together with the corresponding values of $T/T_c$.
For other values of $\beta_3$ used in our calculations we have calculated the
coupling $h$ from Eq. (\ref{h2y}) using the $y$ values from Table III.

\vskip0.5truecm
\begin{center}
\begin{tabular}{|c|c|c|c|c|c|c|c|c|}
\hline 
\multicolumn{1}{|c|}{$x$}& \multicolumn{1}{|c|}{$T/T_c$}  &
\multicolumn{1}{|c|}{$\beta_3$} & \multicolumn{2}{|c|}{I} &
\multicolumn{2}{|c|}{II} & \multicolumn{2}{|c|}{pert} \\
\hline
 &~~&~~& $h$ &$y$ &$h$ & $y$ &$h$ & $y$\\
\cline{4-9}
 ~0.099~&~ 2.0~ &~11~ &~-~ &~-~ &~-~ &~-~ &~ -0.3773~ &~ 0.2274~ \\
 ~0.090~& ~2.8~ &~16~ &~ -0.2611~ &~ 0.3556~ &~-0.2622~ &~ 0.4007~ &~ -0.2700~ &
 ~0.2755~ \\
 ~0.070~&~ 9.0~ &~16~ &~ -0.2528~ &~ 0.4408~ &~-0.2490~ &~ 0.5009~ &~ -0.2588~ &
 ~0.3470~ \\
 ~0.050~& ~72.1~&~16~ &~ -0.2365~ &~ 0.5914~ &~-0.2314~ &~ 0.6721~ &~ -0.2437~ &
 ~0.4756~ \\
 ~0.030~& ~9200~&~16~ &~ -0.2085~ &~ 0.9279~ &~-0.2006~ &~ 1.0544~ &~ -0.2181~ &
 ~0.7758~ \\
\hline
\end{tabular}
\vskip 0.5truecm
\end{center}
\noindent TABLE III . Numerical values for the coupling $h$ and 
continuum mass parameter $y=m_{D0}^2/g_3^2$  of the effective
3d theory in our simulations and the corresponding values of the temperature. 
I and II denote two sets of $h$-values taken from Ref. \cite{karsch98}. The last 
two columns give the coupling $h$ and mass parameter $y$ obtained from the perturbative 
relations Eqs. (\ref{h4d}) and (\ref{y1}). 
\vskip0.5truecm

\begin{figure}
\epsfysize=6cm
\epsfxsize=8cm
\centerline{\epsffile{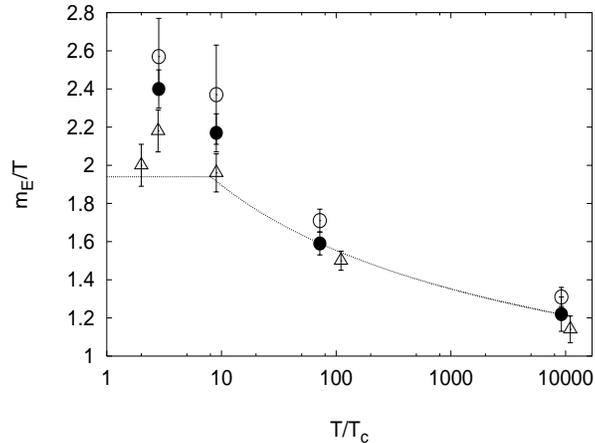}}
\caption{The electric screening masses in units of the temperature.
Shown are the electric masses
$m_E$ for the first (filled circles) and the second (open circles)
set of $h$.
The line represent the fit for the temperature
dependence of the electric mass from 4d
simulations.
The open triangles are the values of the electric mass for $h$ values
obtained from perturbative reduction in the metastable region
(the last column in Table III).
Some data points at the temperature $T \sim 70 T_c$
and $T \sim 9000 T_c$ have been shifted in the temperature scale for
better visibility.
}
\label{md}
\end{figure}

The electric propagators show exponential decay at large distances. 
Therefore it is possible to extract screening masses from them. The procedure
of extracting the screening masses from the propagators was described in 
Ref. \cite{karsch98}. In Figure \ref{md} we compare the electric screening masses
extracted from the Landau gauge propagators in the 3d adjoint Higgs model
calculated on $32^2\times 64$ and $32^2\times 96$ lattices
with the corresponding results from 4d SU(2) gauge theory \cite{heller98}.
The parameters of the effective theory were chosen according to Table III.

As one can see from the figure the agreement between the masses
obtained from 4d and 3d simulation is rather good.
The electric mass shows
some dependence on $h$. For relatively low temperatures ($T<50 T_c$)
the best agreement with the 4d data for the electric mass
is obtained for values of $h$
corresponding to 2-loop dimensional reduction and lying in the
metastable region. This fact motivated our choice of the parameters of
the effective theory at $T=2 T_c$ in section III. 
For higher temperatures, however,
practically no distinction
can be made between the three choices of $h$.

Before closing the discussion on the choice of the parameters of the
effective 3d theory let us compare our procedure of fixing the parameters
of the effective theory with that proposed in \cite{lacock92}.
In Ref. \cite{lacock92} the gauge coupling was fixed by matching the data
on Polyakov loop correlators determined in lattice simulation to the
corresponding value calculated in lattice perturabtion theory. The
resulting gauge couplings turned out to be considerably smaller
than the corresponding ones used in our analysis, e.g. for $T=2 T_c$ it gave
$g^2=1.43$ while our procedure gives $g^2(2 T_c)=2.89$. The scalar 
couplings were fixed according to 1-loop perturbation theory \cite{lacock92}.
Using this procedure the authors of Ref. \cite{lacock92} obtained a good 
description of the
Polyakov loop correlator in terms of the 3d effective theory, however, the spatial
Wilson loop whose large distance behavior determines
the spatial string tension was not well described  in the reduced 3d theory.
The reason for this is the fact that the value of the Polyakov loop
correlator is cut-off dependent and therefore it is not very useful for
extracting the renormalized coupling.


\end{document}